\documentstyle[prl,aps,epsf]{revtex}
\begin{document}
\draft
\twocolumn[\hsize\textwidth\columnwidth\hsize\csname@twocolumnfalse\endcsname
\title{Surface Excitations in a Bose-Einstein Condensate}
\author{R. Onofrio, D.S. Durfee, C. Raman, 
M. K\"{o}hl, C.E. Kuklewicz, and W. Ketterle}
\address{Department of Physics and Research Laboratory of Electronics, \\
 Massachusetts Institute of Technology, Cambridge, MA 02139}
\date{\today{}}
\maketitle
\begin{abstract}
Surface modes in a Bose-Einstein condensate of sodium atoms
have been studied. We observed excitations of standing and rotating 
quadrupolar and  octopolar modes. 
The modes were excited with high spatial and temporal resolution 
using the optical dipole force of a rapidly scanning laser beam. 
This novel technique is very flexible and should be useful for the study of 
rotating Bose-Einstein condensates and vortices.
\end{abstract}
\vskip1pc
]
Elementary excitations play a crucial role in the understanding of 
many-body quantum systems. 
Landau derived the properties of superfluid liquid helium 
from the spectrum of collective excitations \cite{LANDAU}. 
After the observation of Bose-Einstein condensation in dilute alkali 
gases \cite{BEC}, considerable theoretical and experimental efforts focused 
on collective excitations. This has already led to advances in our 
understanding of the weakly interacting Bose gas \cite{DALFOVORMP}. 
In most studies, collective modes were excited by modulating
the parameters of the magnetic trapping potential \cite{JIN1,MEWES}. 
This method of exciting collective modes is limited to spatial 
perturbations that reflect the geometry of the trapping coils. 
Such a limitation is particularly severe for the widely used dc 
magnetic traps, where only modes with
cylindrical symmetry have been excited \cite{NOTE1}. 

Studies of high multipolarity modes are important for a number of reasons.
First, high multipolarity modes are the closest
counterpart to the surface excitations in mesoscopic liquid helium droplets.
These surface modes are considered crucial to understand finite size effects
in superfluids, but are difficult to achieve experimentally \cite{CHIN}.
Second, for higher angular momentum the surface modes change their 
character from collective to single particle type \cite{DALFOVORMP}. 
This crossover could be crucial for the existence of a critical rotational 
velocity for vortex formation \cite{DALFOVO,LUNDH}. 
Also, because the thermal atoms are localized around the
Thomas-Fermi radius, surface modes should be more sensitive to 
finite temperature effects \cite{DALFOVO}. 

In this Letter we report on the observation of surface excitations 
of a Bose-Einstein condensate confined in a dc magnetic trap. 
The excitations were induced by the optical dipole force of  
a focused red-detuned laser beam which was controlled by a 
2-axis acousto-optic deflector. 
With these tools, local and controllable deformations of 
the magnetic trapping potential with both arbitrary spatial 
symmetry and timing can be achieved. This opens the way
to selectively excite modes with higher multipolarity and 
complex spatial patterns.

Elementary excitations in a dilute Bose condensate are usually 
described by the hydrodynamic equations derived from the 
Bogoliubov theory \cite{BOGOLIUBOV}, 
which closely resemble the equations describing superfluids 
at zero temperature \cite{PINES}:
\begin{equation}
m{\partial \over \partial t}{\bf v}+\nabla
\left( {1\over 2}mv^{2}+V_{ext}({\bf r})-\mu +{4\pi \hbar ^{2}a\over m}\rho \right) =0.
\end{equation}
 Here \( \rho ({\bf r},t) \) and \( {\bf v}({\bf r},t) \) 
are the condensate density and velocity respectively 
(linked by a continuity equation), $m$ the atomic mass, 
\( a \) the $s$-wave scattering length, \( \mu  \) the chemical
potential, and \( V_{ext} \) the external trapping potential. 
For an isotropic harmonic oscillator potential 
\( V_{ext}=m \omega_{0}^{2}r^{2}/2 \) 
the solution
for the density perturbation \( \delta \rho  \) can be expressed as:
\begin{equation}
\delta \rho ({\bf r})=P_{\ell }^{(2n)}(r/R)r^{\ell }Y_{\ell m}(\theta ,\phi ),
\end{equation}
 where \( P_{\ell }^{(2n)}(r/R) \) are polynomials of degree \( 2n \) 
(\( R \) being the Thomas-Fermi radius 
\( R=\sqrt{2\mu /m \omega_0^{2}} \), 
\( Y_{\ell m}(\theta ,\phi ) \) are the spherical harmonics, and \( \ell  \), 
\( m \) are the total angular momentum of the excitation and 
its \( z \) component, respectively. 
The dispersion law for the frequency of the normal modes is 
expressed in terms of the trapping frequency \( \nu_0=\omega_0/2 \pi \) as 
\cite{STRINGARI}:
\begin{equation}
\nu (n,\ell )=(2n^{2}+2n\ell +3n+\ell )^{1/2}\nu _{0}, 
\end{equation}
 which should be compared to the prediction for an ideal Bose gas
in a harmonic trap, \( \nu _{HO}=(2n+\ell )\nu _{0} \). The
effect of interactions in determining the transition from a collective to a
single-particle regime is particularly evident for the excitations whose radial
dependence of the density perturbation has no nodes ($n$=0).
These modes are  referred to as {\it surface excitations} since the density
perturbation, while vanishing at the origin, is peaked at the surface of 
the condensate. In thin films of superfluid liquid ${}^4$He and ${}^3$He, 
their study has led to the observation of third sound \cite{THIRDSOUND}.
In a semiclassical picture these excitations can be considered the 
mesoscopic counterpart to tidal waves at the
macroscopic level \cite{NEWTON,ALKHAWAJA}.

The experimental results were obtained using a newly developed apparatus
for studying Bose-Einstein condensates of sodium atoms. 
A Zeeman slower with magnetic field reversal \cite{WITTE}
delivers \( 10^{11} \) slow atoms s\( ^{-1} \) which
are collected in a magneto-optical (MOT) trap. A loading time of 3 s allowed 
us to obtain \( 10^{10}-10^{11} \) atoms in a dark-SPOT trap \cite{KETTERLE}
at \( \simeq 1 \) mK. After 5 ms of polarization gradient cooling, atoms in
the \( F=1,m_{F}=-1 \) ground state at \( \simeq 50-100\mu  \)K are loaded into
a magnetic trap. 
The latter realizes a Ioffe-Pritchard configuration modified with 
four Ioffe bars and two strongly elongated pinch coils, symmetrically 
located around a quartz glass cell.
This novel design combines excellent optical access with tight confinement. 
The typical values for the axial curvature and radial gradient
of the magnetic field at the trap center are \( 202 \) G/cm\( {}^{2} \) 
and \( 330 \) G/cm, among the largest ever obtained in such magnetic traps.
The resulting trapping frequencies are \( \nu _{r}=547 \) Hz and
\( \nu _{z}=26 \) Hz for the radial and the axial directions respectively.
The background gas-limited lifetime of the atoms in the magnetic trap at
\( \simeq 10^{-11} \) Torr is around 1 min.
After evaporative cooling with an rf-sweep lasting 20 s, around
\( 5-10 \) million atoms are left in a condensate with a chemical potential of
\( 200 \) nK and a negligible thermal component (condensate fraction \( \geq 90\% \)).
A decompression to the final radial and axial trapping frequencies of
\( (90.1\pm 0.5) \)
Hz and \( 18 \) Hz lowered the density to \( 2\cdot 10^{14} \)
cm\( {}^{-3} \) where three-body recombination losses were less prominent. 
The radial trapping frequency was measured by exciting the condensate 
motion with a short modulation of the bias magnetic field and 
looking at the free center of mass oscillation in the magnetic trap.

Surface modes were excited by perturbing the magnetic trapping potential with
light from a Nd:YAG laser (emitting at 1064 nm) travelling parallel to the
axis of the trap and focused near the center of the magnetic trap.
Because of the low intensity of the laser beam and the large detuning from the
sodium resonance, heating from spontaneous scattering was negligible.
The laser beam was red-detuned from the sodium resonance and therefore gave
rise to an attractive dipole potential \cite{CHU}.
The 1 mm Rayleigh range of the beam waist is considerably longer than the 
220 \( \mu \)m axial extent of the condensate. 
Therefore, the laser only created radial inhomogeneities in the 
trapping potential, leaving the axial motion almost undisturbed. 

The spatial and temporal control
of the beam was achieved with two crossed acousto-optic deflectors. Using the 2-axis
deflector arbitrary laser patterns could be scanned in a plane transverse
to the propagation of the laser beam. The maximum size of these
patterns is 100 beam widths in both directions. The scan rate was chosen to
be 10 kHz, which is much larger than the trapping frequencies. Thus, the atoms
experienced a time-averaged potential that is superimposed upon the magnetic
trap potential as depicted in Fig. 1a. For these experiments a beam width of
15 \( \mu  \)m and a power of \( 80 \mu  \)W were used to generate a potential depth
corresponding to 20 \( \% \) of the chemical potential for each point.

\begin{figure}
\epsfxsize=70mm
\centerline{\epsfbox{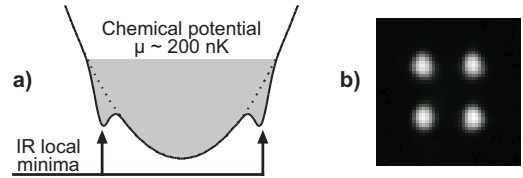}}
\vspace{0.5cm}
\caption{
Scheme for exciting collective modes by optically deforming the 
magnetic trap potential.
(a) The combined potential of the magnetic trap and two red-detuned beams.
(b) An image of the four-point IR pattern taken with a CCD camera.
The laser beams were arranged on the corners of a 
40 \protect\( \mu \protect \)m \protect\( \times \protect \)
40 \protect\( \mu \protect \)m square centered on a condensate
with a Thomas-Fermi radius of
\protect\( \simeq \protect \) 25 \protect\( \mu \protect \)m. 
}
\end{figure}

For an anisotropic axially symmetric trapping potential, only the \( z \) 
component of the angular momentum is conserved and the eigenfunctions are more 
complicated than in the isotropic case. However, surface modes of the 
form as in Eq. (2) with \( m=\pm \ell  \) are still solutions with a 
frequency \cite{STRINGARI}:
\begin{equation}
\label{MODEASYM}
\nu (m=\pm \ell )=\sqrt{\ell }~\nu _{r}.
\end{equation}

Quadrupolar standing waves were studied by exciting a superposition of
\( \ell =2,m=2 \) and \( \ell =2,m=-2 \) modes
with a pattern of two points located on opposite sides of the condensate.
The light intensity was modulated in phase at the expected quadrupole frequency
\( \nu _{2}=\sqrt{2}\nu _{r} \). After 5 cycles the IR light was turned off,
leaving the condensate free to oscillate in the magnetic trap. The condensate
was then released from the magnetic trap and after 20 ms of ballistic expansion
it was probed by resonant absorption imaging along the axis of the trap.
In Fig. 2a, images are shown for different phases of the oscillation.
The aspect ratio of the condensate oscillates at a frequency of 
\( (130.5\pm 2.5) \) Hz with a damping time of about 0.5 s.  
A similar damping time was observed for the lowest \( m=0 \) mode 
of an almost pure condensate \cite{STAMPER}. 

A rotating wave  \( \ell =2,m=2 \) was excited  with two 
IR spots of constant intensity rotating around the axis at 
half the measured quadrupole frequency. This excitation scheme
was highly frequency selective. When the rotation frequency deviated by 
\( 10\% \) from the resonance no excitation of this mode was
observed, consistent with the narrow bandwidth of the mode. 
In Fig. 2b we show a set of 10 pictures of the rotating mode
taken with non-destructive phase-contrast imaging
\cite{PHASECONTRAST} in the magnetic trap.

The higher lying \( \ell  \)=4 surface mode (superposition of \( m=\pm 4 \))
was driven with a four-point pattern that was intensity-modulated at 
the expected frequency \( \nu _{4}=2\nu _r \) (Fig. 1b). 
In Fig. 3a time of flight absorption images
are shown for variable hold times in the magnetic trap after stopping 
the drive and compared to the time evolution for a pure \( \ell  \)=4 surface
mode (Fig. 3b). By analyzing the density distribution close to the surface
we extracted the Fourier spectrum of the first radial moment 
\( r=r(\theta ) \), which was strongly peaked at \( \ell =4 \).
The time evolution of the \( \ell =4 \)
Fourier cosine coefficient was obtained by repeating the analysis for
various hold times (Fig. 3c).

\begin{figure}
\epsfxsize=80mm
\centerline{\epsfbox{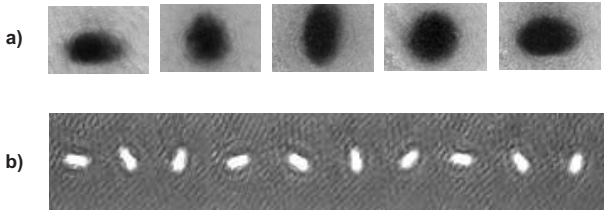}}
\vspace{0.5cm}
\caption{Observation of standing and rotating surface waves.
(a) shows absorption images of a standing 
\protect\( \ell =2,m=\pm 2\protect \)
quadrupole mode. They  were taken after  free oscillation times  of 3.5, 5.25,
7, 8.75 and 10.5 ms (from left to right) in the magnetic trap and 20 ms of ballistic expansion.
In (b) multiple phase-contrast images of a clockwise rotating \protect\( \ell =2,m=2\protect \)
quadrupole surface excitation are shown, each frame being 2 ms apart.
The field of view of each image is $720 \mu$m $\times~ 480 \mu$m.}
\end{figure}

\begin{figure}
\epsfxsize=85mm
\centerline{\epsfbox{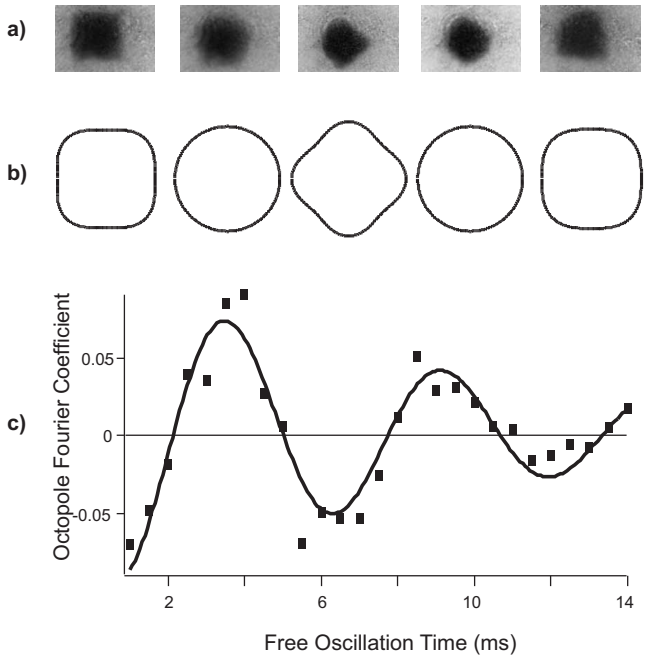}}
\vspace{0.5cm}
\caption{Observation of an octopolar mode of a Bose-Einstein condensate.
Absorption images of a condensate for various hold times (1, 2, 3.5, 4.5 
and 6.5 ms from left to right) in the magnetic trap (a) after the  excitation. 
The shape of a pure \protect\( \ell =4\protect \)
oscillation is schematically depicted in (b) for one cycle.
(c) shows the \protect\( \ell =4\protect \) cosine Fourier coefficient 
as a function of the free oscillation time in the magnetic trap.
The field of view of each image is $720 \mu$m $\times~ 480 \mu$m.}
\end{figure}

We observed an exponentially decaying oscillation at 
\( \nu _{4}=(177\pm 5) \) Hz with a damping time of 
\( \tau _{4}=(9.5\pm 2.2) \)~ ms. The agreement between the 
measured frequencies and the hydrodynamic predictions is very good 
(see Table I). Note that the octopolar mode is damped much faster 
than the quadrupole mode.
This could indicate that higher order surface excitations interact 
more strongly with the thermal cloud.
Due to the mean-field repulsion of the condensate, the effective potential
felt by the thermal atoms has a minimum at the Thomas-Fermi radius.
For increasing \( \ell  \) surface waves are more localized in the
same region (Eq.(2)). Thus, a systematic study of the temperature dependence
of frequencies and damping times of higher order surface modes could extend  
thermometry for Bose-Einstein condensates to lower temperatures where 
no thermal cloud is discernible (and the usual method of fitting the 
wings of the thermal distribution is no longer applicable).

\vspace{0.3cm}

\begin{center}
\begin{tabular}{|l|r|r|r|}
\hline
~~~~\( \ell  \) &
 \( \nu _{\ell } \)~(Hz) ~~~~&
 \( \nu_{\ell } /\nu_1 (exp) \) ~~~~~&
 \( \nu_{\ell } /\nu_1 (th) \) \\
\hline
~~~~1~&
~~~~~90.1 \( \pm \) 0.5 ~~~~&
 \(-\)~~~~~~~~~~ &
 \(-\)~~~~~~\\
 ~~~~2~~~~&
~~~ 130.5 \( \pm  \) 2.5 ~~~~&
~~ 1.45 \( \pm  \) 0.04~~~&
$ \sqrt{2}$~~~ \\
 ~~~~4~~~~&
177 \( \pm  \)5 ~~~~&
~~~1.96 \( \pm  \) 0.06~~~&
~~~2~~~~ \\
\hline
\end{tabular}

\end{center}
Table I: Comparison between observed and predicted frequencies 
for the quadrupole and the octopole surface excitations, normalized 
to the radial trapping frequency (dipole mode) $\nu_1$.
\vspace{0.3cm} 

For higher \( \ell \),  
the crossover from the hydrodynamic regime to the single 
particle picture could be explored. This is  expected to occur
for \( \ell \geq \ell _{crit}=2^{1/3}(R/a_{HO})^{4/3}\simeq 24 \)
\cite{DALFOVO} 
(where \( a_{HO}=[\hbar /2 \pi m {(\nu_r^2 \nu_z)}^{1/3}]^{1/2} \) 
is the harmonic oscillator length), for our trap parameters.
The excitation of such modes would require smaller beam waists. 
However, this resulted either in a very weak excitation or, when the 
power was increased, the condensate became strongly distorted and/or 
localized around the laser focus leading to high densities and 
to large recombination losses. 
We plan to use blue-detuned light in the future, which will make it 
easier to create stronger perturbations without loss mechanisms.

\begin{figure}
\epsfxsize=80mm
\centerline{\epsfbox{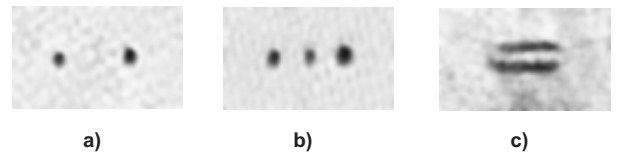}}
\vspace{0.5cm}
\caption{Bose-Einstein condensates in time-averaged potential optical traps. 
Patterns
with two (a) and three (b) points and a double sheet (c) are shown. 
The absorption images are taken along the axis of the laser beam, 
with resonant probe light.
The frame size is
\protect\( 200\mu \protect \)m \protect\( \times \protect \)
\protect\( 120\mu \protect \)m.}
\end{figure}

Our method of generating time-averaged optical potentials can also 
be used to create purely optical traps in a variety of geometries. 
By increasing the laser intensity and shutting
off the magnetic trap we were able to tranfer the condensate into 
multiple optical dipole traps, as shown in Fig. 4.
They can be used for interference of multiple condensates and studies of 
coherence and decoherence. 

Another interesting possibility is the study 
of condensates in rotating potentials where vortices should be stable. 
Our first attempts showed a very short trapping time, probably caused 
by heating due to micromotion. It should be possible to overcome this 
limitation by increasing 
the scan frequency beyond the current maximum value of $100$ kHz. 

In conclusion, we have developed a technique to excite surface modes in
a Bose-Einstein condensate by inducing deformations of the trap potential
with a rapidly scanning red-detuned laser beam. 
With this technique we could excite both standing and rotating modes.  
The measured frequencies for quadrupole and octopole modes are in 
agreement with the predictions of the hydrodynamic theory for collective 
excitations of dilute Bose gases. This flexible technique should be 
useful for the investigation of the interplay between collective excitations 
and the physics of rotating Bose-Einstein condensates.

We would like to thank J. Gore, Z. Hadzibabic, and J. Vogels for experimental
assistance and useful discussions.
This work was supported by the ONR, NSF, JSEP (ARO), NASA, 
and the David and Lucile Packard Foundation. 
M.K. acknowledges also support from Studienstiftung des Deutschen Volkes.

\end{document}